# Artificial Intelligence-Based Image Enhancement in PET Imaging: Noise Reduction and Resolution Enhancement

Juan Liu, PhD[1,†], Masoud Malekzadeh, MS[2,†], Niloufar Mirian, MS[1], Tzu-An Song, MS[2], Chi Liu, PhD[1,*], Joyita Dutta, PhD[2,3,*]

[1] Department of Radiology and Biomedical Imaging, Yale School of Medicine, New Haven, CT, USA

[2] Department of Electrical and Computer Engineering, University of Massachusetts Lowell, Lowell, MA, USA

[3] Gordon Center for Medical Imaging, Massachusetts General Hospital and Harvard Medical School, Boston, MA, USA

## Abstract

High noise and low spatial resolution are two key confounding factors that limit the qualitative and quantitative accuracy of PET images. AI models for image denoising and deblurring are becoming increasingly popular for post-reconstruction enhancement of PET images. We present here a detailed review of recent efforts for AI-based PET image enhancement with a focus on network architectures, data types, loss functions, and evaluation metrics. We also highlight emerging areas in this field that are quickly gaining popularity, identify barriers to large-scale adoption of AI models for PET image enhancement, and discuss future directions.

## 1. Introduction

Positron emission tomography (PET) is a noninvasive molecular imaging modality which is increasingly popular in oncology, neurology, cardiology, and other fields [1-3]. Accurate quantitation of PET radiotracer uptake is vital for disease diagnosis, prognosis, staging, and treatment evaluation. The key confounding factors compromising PET image quality and quantitative accuracy are high noise and low spatial resolution. The high noise issue could be further accentuated in scenarios where the radiotracer dose is reduced in order to decrease a patient's radiation exposure or the scan time is reduced, possibly with the intent of increasing throughput or decreasing patient discomfort. A variety of physical and algorithmic factors contribute to the resolution limitations of PET, including the non-collinearity of the emitted photon pairs, finite crystal size, crystal penetration and inter-crystal scatter of the detected photons, positron range, and within- or post-reconstruction smoothing. The low resolution of PET images manifests as partial volume effects, which lead to the spillover of estimated activity across different regions-of-interest (ROIs). Collectively, the high noise and low resolution in PET images could lead to over- or under-estimation of standardized uptake values (SUVs) in an ROI and inaccurate lesion delineation or detection.

---

[†] Equal contribution

[*] Co-corresponding authors. Emails: dutta.joyita@mgh.harvard.edu and chi.liu@yale.edu

A variety of within- and post-reconstruction approaches have been developed to denoise and/or deblur PET images. Among within-reconstruction approaches, regularization using a quadratic penalty and early iteration termination are some of the most rudimentary techniques to tackle image noise [4-6], while point spread function (PSF) modeling in the image or sinogram domain is a common approach to improve resolution [7, 8]. Many penalized likelihood reconstruction approaches have been proposed to improve the noise and resolution characteristics of the reconstructed images [9]. Among post-reconstruction PET image enhancement approaches, Gaussian filtering is the most commonly used denoising technique and is built into the reconstruction software for most clinical and small animal PET scanners. Given the resolution loss associated with Gaussian filters, many edge-preserving alternatives have been proposed for PET image denoising. These include bilateral [10, 11], anisotropic diffusion [12, 13], wavelet-domain [14-17], frequency-domain [18], and nonlocal [19-21] filters. While several penalized deconvolution approaches have been proposed for resolution enhancement [22, 23], the most popular resolution recovery approaches in PET are simple partial volume correction techniques that rely on a segmented anatomical template [24-26]. Some denoising approaches and nearly all deblurring approaches utilize high-resolution anatomical information based on structural magnetic resonance imaging (MRI) or computed tomography (CT).

With the advent of a new era in computing largely dominated by deep learning and artificial intelligence (AI), the fields of medical image reconstruction and processing have been revolutionized by the availability of a new generation of AI-powered image enhancement tools. This review will focus on AI-based approaches for PET image enhancement that have been developed in the recent years and will also discuss new developments and emerging technologies in this arena. Because AI-based reconstruction techniques are the subject of Kuang Gong and colleagues' article, "The Evolution of Image Reconstruction in PET: From Filtered Back-Projection to Artificial Intelligence," in this Special Issue, this article focuses exclusively on AI-based postreconstruction PET image enhancement. We should note here that, while within-reconstruction image enhancement techniques tend to have improved bias-variance characteristics compared to their post-reconstruction counterparts, the latter category of methods is attractive due to two main reasons. Firstly, since these techniques do not require access to raw sinogram or list-mode data, they cater to a wider base of users who only have access to the final reconstructed images. Secondly, data size is a major factor affecting the accuracy of AI models and most large image repositories and databases contain only the reconstructed images but no raw data, making it more practical to train AI models that receive image-domain inputs.

It should be noted that most large data repositories that enable robust training and validation of AI-based image enhancement models are acquired over multiple hospitals and imaging centers. For these multi-site studies, inter-site differences in scanner models and image acquisition protocols (including radiotracer dosage, scan duration, and image reconstruction parameters) could lead to significant variability in image characteristics. As a result, many image enhancement techniques are motivated by the need for data harmonization across multiple data sources so as to enable collective use. This review will discuss emerging PET image enhancement techniques that are particularly useful for data harmonization. A major bottleneck toward the robust training and validation of AI models for medical imaging modalities like PET is the need for centralized curation of data, which is mired in patient privacy regulations and data ownership concerns. Federated learning, which involves the sharing of deep learning models across institutions in place of data sharing, offers a promising path to circumvent the data sharing bottleneck and facilitate AI practice in the medical domain. We will also discuss emerging trends in federated learning as it relates to PET.

In section 2, we tabulate and discuss recently published AI-based PET image enhancement approaches, including both supervised and unsupervised techniques. In section 3, we highlight several active areas of research that are of direct relevance to PET image enhancement. Finally, in section 4, we discuss future directions, identify barriers, and conclude the paper.

## 2. Summary of Existing Methods

### 2.1. Overview

The PET image denoising and deblurring problems both involve the retrieval or restoration of a clean image from its corrupt counterpart. For the denoising problem, the corrupt image is a low-count (also referred to as low-dose) image while the clean image is a higher-dose image. For the deblurring problem, the corrupt image is a lower-resolution counterpart of a higher-resolution clean image. Unlike model-based approaches employed in conventional image processing where the denoising and deblurring problems are associated with very different models, the difference between AI-based image deblurring and denoising often lies in the data used for training. It should be noted here that, while a methodological dichotomy exists, in many practical scenarios, the objective may be to simultaneously deblur and denoise an image.

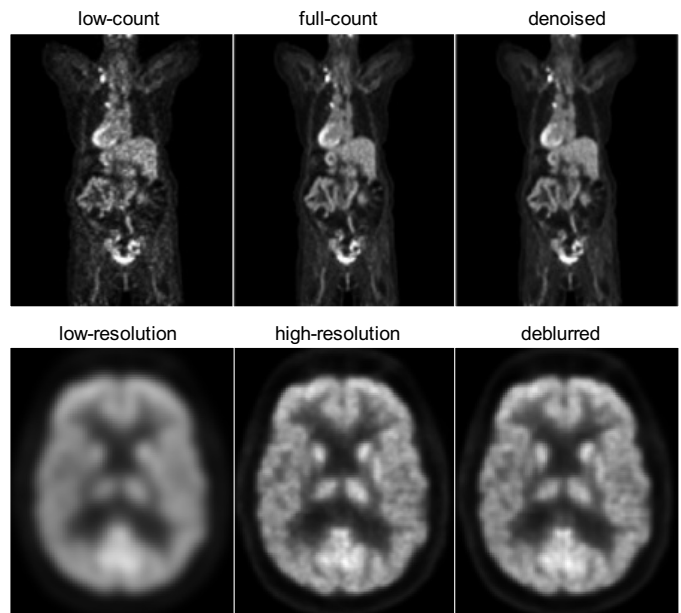

**Fig. 1.** Examples of corrupt (left), clean (middle), and enhanced (right) images for the denoising problem (top) and the deblurring problem (bottom). In the denoising example, a 3D U-Net network was used to produce the denoised image from a 20% low-count image. In the deblurring example, a very deep convolutional neural network was used to generate

a high-resolution PET image similar to those produced by the Siemens HRRT dedicated brain scanner from a low-resolution one generated by the Siemens HR+.

Fig. 1 shows examples of corrupt, clean, and restored/enhanced images for the denoising and deblurring problems. The images are all based on human PET imaging using the $^{18}$F-fluorodeoxyglucose ($^{18}$F-FDG) radiotracer. For both the denoising and deblurring examples, the enhanced images were produced by neural networks that received the corrupt image as its input.

The denoising and deblurring problems aim to compute an estimate, $\hat{x}$, of a clean image $x$ from its corrupt counterpart $y$ in the form of some explicit or implicit function, $\hat{x} = \mathcal{R}(y)$. Traditional image processing approaches compute the unknown image using either an analytic formula or a model-based iterative procedure. Unlike traditional analytic or iterative image processing methods, AI-based approaches for image enhancement *learn* the mapping from the corrupt to the clean image directly from the data. For a parametric model, this mapping is of the form $\hat{x} = f(y, w)$, where the vector $w$ is a set of "weight" parameters for the neural network computed during the network training phase. This is achieved by minimizing a loss function. The majority of AI models for image enhancement are based on *supervised* learning. A supervised learning setup requires paired clean and corrupt images for network training, as illustrated in Fig. 2. Following the training phase, the model is tested or validated on an independent subset of the data by comparing the enhanced images with the clean (ground truth) images using one or more evaluation metrics. While supervised models are simpler to design and train, their need for paired clean and corrupt inputs limits their applicability to cases where clean counterparts of corrupt images may not be available. In contrast, *unsupervised* learning models obviate the need for paired inputs and are currently an active area of investigation for PET image enhancement.

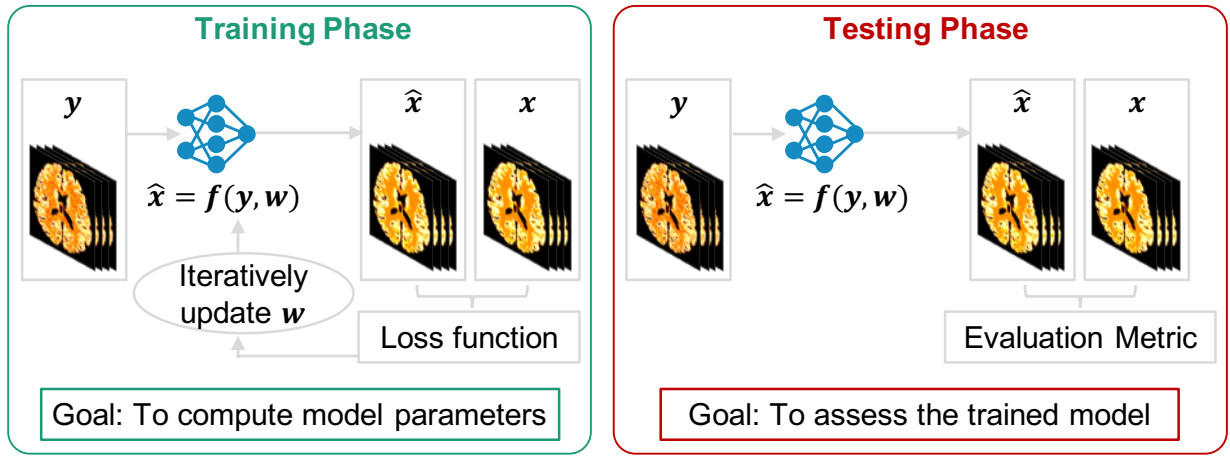

**Fig. 2**. A typical supervised learning setup for image enhancement with a training phase that computes the model parameters and requires paired clean and corrupt images and a testing phase that assesses the model's performance in an independent subset of paired images.

Table 1 presents a survey and summary of some noteworthy AI-based image enhancement efforts for PET. In the following subsections, we discuss the supervised and unsupervised model architectures, loss functions, and evaluation metrics for several of the tabulated methods.

Table 1. A summary of deep learning techniques for PET image enhancement

| Task | Learning Style | Paper | Method and Architecture | Data and Radiotracer | Loss Function | Input and Output | Evaluation Metric |
|---|---|---|---|---|---|---|---|
| Denoising | Supervised | da Costa-Luis et al. [27, 28] | 3D CNN, 3 layers | $^{18}$F-FDG simulation and human data | L2 loss | Input: low-count PET images with and without resolution modeling, T1-weighted MR imaging, and T1-guided NLM filtering of the resolution modeling reconstruction. Output/Training target: full-count PET | NRMSE, Bias vs. variance curves |
| | | Gong et al. [29] | CNN with residual learning, 5 residual blocks | $^{18}$F-FDG simulation data, $^{18}$F-FDG human data | L2 loss + perceptual loss | Input: low-count PET. Output/Training target: full-count PET | CRC vs. variance curves |
| | | Xiang et al. [30] | Deep auto-context CNN, 12 convolutional layers | $^{18}$F-FDG human data | L2 loss + L2 norm weight regularization | Input: low-count PET, T1-weighted MR imaging. Output/Training target: full-count PET | NRMSE, PSNR |
| | | Chen et al. [31] | 2D residual U-Net | $^{18}$F-Florbetaben human brain data | L1 loss | Input: low-count PET, multi-contrast MR imaging (T1-weighted, T2-weighted, T2 FLAIR). Output/Training target: full-count PET | NRMSE, PSNR, SSIM |
| | | Spuhler et al. [32] | 2D residual dilated CNN | $^{18}$F-FDG human data | L1 loss | Input: low-count PET. Output/Training target: full-count PET | SSIM, PSNR, MAPE |
| | | Serreno-Sosa et al. [33] | 2.5D U-Net with residual learning and dilated convolution | $^{18}$F-FDG human brain data | - | Input: low-count PET. Output/Training target: full-count PET | SSIM, PSNR, MAPE |
| | | Schaefferkoetter et al. [34] | 3D U-Net | $^{18}$F-FDG human data | L2 loss | Input: low-count PET. Output/Training target: full-count PET | CRC |

| | | | | | | |
|---|---|---|---|---|---|---|
| | | **Sano et al.** [35] | **2D residual U-Net** | Proton-induced PET data from simulations and a human head and neck phantom study | L2 loss | Input: noisier low-count PET Output/Training target: less noisy low-count PET | PSNR |
| | | **Wang et al.** [36] | **GAN** Generator: 3D U-Net Discriminator: 4-convolution layer CNN | $^{18}$F-FDG simulated data, $^{18}$F-FDG human brain data | L1 loss + adversarial loss | Input: low-count PET, T1-weighted MR imaging, fractional anisotropy and mean diffusivity images computed from diffusion MR imaging Output/Training target: full-count PET | PSNR, SSIM |
| | | **Zhao et al.** [37] | **CycleGAN** Generator: multi-layer CNN Discriminator: 4-convolution layer CNN | $^{18}$F-FDG simulated data, $^{18}$F-FDG human data | L1 supervised loss + Wasserstein adversarial loss + cycle-consistency loss + identity loss | Input: low-count PET Output/Training target: full-count PET | NRMSE, SSIM, PSNR, learned perceptual image patch similarity, SUV bias |
| | | **Xue et al.** [38] | **Least squares GAN** Generator: 3D U-Net like network with residual learning and self-attention modules Discriminator: 4-convolution layer CNN | $^{18}$F-FDG human data | L2 loss + adversarial loss | Input: low-count PET Output/Training target: full-count PET | PSNR, SSIM |
| | | **Wang et al.** [39] | **cGANs with progressive refinement** Generator: 3D U-Net Discriminator: 4-convolution layer CNN | $^{18}$F-FDG human brain data | L1 supervised loss + adversarial loss | Input: low-count PET Output/Training target: full-count PET | NMSE, PSNR, SUV bias |
| | | **Kaplan et al.** [40] | **GAN** Generator: 2D encoder-decoder with skip connection Discriminator: 5-layer CNN | $^{18}$F-FDG human data | L2 loss + gradient loss + total variation loss + adversarial loss | low-count PET Output/Training target: full-count PET | RMSE, MSSIM, PSNR |
| | | **Zhou et al.** [41] | **CycleGAN** Generator: multi-layer 2D CNN Discriminator: 6-layer CNN | $^{18}$F-FDG human data | L1 supervised loss + Wasserstein adversarial loss + cycle-consistency loss + identity loss | Input: low-count PET Output/Training target: full-count PET | NRMSE SSIM PSNR SUV bias |
| | | **Ouyang et al.** [42] | **GAN** Generator: 2.5D U-Net Discriminator: 4-convolution layer CNN | $^{18}$F-florbetaben human data | L1 loss + adversarial loss + task-specific perceptual loss | Input: low-count PET Output/Training target: full-count PET | SSIM PSNR RMSE |

| Author | Method | Data | Loss | Input/Output | Metrics |
|---|---|---|---|---|---|
| Gong et al. [43] | **GAN** Generator: hybrid 2D and 3D encoder-decoder Discriminator: 6-layer CNN | $^{18}$F-FDG human data | L2 loss + Wasserstein adversarial loss | Input: low-count PET Output/Training target: full-count PET | NRMSE, PSNR, Riesz transform-based feature similarity index, visual information fidelity |
| Liu et al. [44] | **3D U-Net** cross-tracer cross-protocol transfer learning | $^{18}$F-FDG human data, $^{18}$F-FMISO human data, $^{68}$Ga-DOTATATE data | L2 loss | Input: low-count PET Output/Training target: full-count PET | NRMSE, SNR, SUV bias |
| Lu et al. [45] | Network comparison: Convolutional autoencoder, U-Net, residual U-Net, GAN, 2D vs. 2.5D vs. 3D | $^{18}$F-FDG human lung data | L2 loss | Input: low-count PET Output/Training target: full-count PET | NMSE, SNR, SUV bias |
| Ladefoged et al. [46] | **3D U-Net** | $^{18}$F-FDG human cardiac data | Huber loss | Input: low-count PET, CT Output/Training target: full-count PET | NRMSE, PSNR, SUV bias, |
| Sanaat et al. [47] | **3D U-Net** | $^{18}$F-FDG human data | L2 loss | Input: low-dose PET image/sinogram Output/Training target: standard-dose PET image/sinogram | RMSE, PSNR, SSIM, SUV bias |
| He et al. [48] | **Deep CNN** | $^{18}$F-FDG simulated brain data, $^{18}$F-FDG dynamic data | L1 loss + gradient loss + total variation loss | Input: noisy dynamic PET image, MR imaging Output/Training target: composite dynamic images | RMSE, SSIM, CRC vs. variance curves |
| Wang et al. [49] | **Deep CNN** | $^{18}$F-FDG human whole-body data | Attention-weighted loss | Input: low-count PET, T1-weighted LAVA MR imaging Output/Training target: full-count PET | NRMSE, SSIM, PSNR, SUV bias |
| Schramm et al. [50] | **3D CNN** with residual learning | $^{18}$F-FDG, $^{18}$F-PE2I, $^{18}$F-FET human data | L2 loss | Input: OSEM-reconstructed low-count PET, T1-weighted MR imaging Output/Training target: enhanced PET (based on anatomical guidance) | CRC, SSIM |
| Jeong et al. [51] | **GAN** Generator: 2D U-Net Discriminator: 3-layer CNN | $^{18}$F-FDG human brain data | L2 loss + adversarial loss | Input: low-count PET Output/Training target: full-count PET | NRMSE, PSNR, SSIM, SUV bias |

| | | Tsuchiya et al. [52] | **2D CNN** with residual learning | $^{18}$F-FDG human whole-body data | Weighted L2 loss | Input: low-count PET, Output/Training target: full-count PET | SUV bias |
|---|---|---|---|---|---|---|---|
| | | Liu et al. [53] | **2D U-Net** with asymmetric skip connections | Simulated $^{18}$F-FDG brain data | L2 loss | Input: filtered backprojection reconstructed PET, T1-weighted MR imaging Output/Training target: MLEM-reconstructed PET | MSE, CNR, bias-variance images |
| | | Sanaat et al. [54] | **CycleGAN** Generator: 2D U-Net like network Discriminator: 9-layer CNN **ResNet** 20 convolutional layers | $^{18}$F-FDG human data | CycleGAN: L1 loss + adversarial loss ResNet: L2 loss | Input: low-count PET Output/Training target: full-count PET | MSE, PSNR, SSIM, SUV bias |
| | | Chen et al. [55] | **2D U-Net** with residual learning | $^{18}$F-FDG human brain data | L1 loss | Input: low-count PET, multi-contrast MR imaging (T1-weighted, T2-weighted, T2 FLAIR) Output/Training target: full-count PET | RMSE, PSNR, SSIM |
| | | Katsari et al. [56] | **SubtlePET™ AI** | $^{18}$F-FDG PET/CT human data | - | - | SUV bias, Subjective image quality, lesion detectivity |
| | Unsupervised, weakly-supervised, or self-supervised | Cui et al. [57] | **Deep Image Prior** 3D U-Net | Simulation and human data from two radiotracers: Ga-PRGD2 (PET/CT) and $^{18}$F-FDG (PET/MR) | L2 loss | Inputs: CT/MR image Output: denoised PET Training target: noisy PET | CRC vs. variance curves |
| | | Hashimoto et al. [58] | **Deep Image Prior** 3D U-Net | $^{18}$F-FDG simulated data, $^{18}$F-FDG monkey data | L2 loss | Input: static PET image Training target: noisy dynamic PET image Output: denoised dynamic PET image | PSNR, SSIM, CNR |
| | | Hashimoto et al. [59] | **4D Deep Image Prior** Shared 3D U-Net as feature extractor and reconstruction branch for each output frame | $^{18}$F-FDG simulated data and $^{18}$F-FDG and $^{11}$C-raclopride monkey data | Weighted L2 loss | Input: static PET image Training target: 4D dynamic PET image Output: denoised dynamic PET image | Bias vs. variance curves, PSNR, SSIM |

| | | Wu et al. [60] | **Noise2Noise** 3D CNN encoder-decoder | $^{15}$O-water human data | L2 denoising loss + L2 bias control loss + L2 content loss | Inputs: low-count PET images from one injection Output: denoised low-count PET Training target: low-count PET images from another injection | CRC |
|---|---|---|---|---|---|---|---|
| | | Yie et al. [61] | **Noisier2Noise** 3D U-net | $^{18}$F-FDG human data | L2 loss | Inputs: extreme low-count PET Output: denoised low-count PET Training target: low-count PET | PSNR, SSIM |
| Deblurring | Supervised | Song et al. [62] | **Very Deep CNN** 20-layer CNN with residual learning | $^{18}$F-FDG simulation and human data | L1 loss | Input: low-resolution PET, T1-weighted MR imaging, spatial (radial + axial) coordinates Output/Training target: high-resolution PET | PSNR, SSIM |
| | | Gharedaghi et al. [63] | **Very Deep CNN** 16-layer CNN with residual learning | Human data, radiotracer unknown | L2 loss | Input: low-resolution PET, Output/Training target: high-resolution PET | PSNR, SSIM |
| | | Chen et al. [64] | **CycleGAN** Model trained on simulation data and applied to clinical data | $^{18}$F-FDG simulated images for training and human images for validation | Adversarial loss + cycle-consistency loss | Input: low-resolution PET, Output/Training target: high-resolution PET | Visual examples only, no quantitative results |
| | Unsupervised, weakly-supervised, or self-supervised | Song et al. [65] | **Dual GANs** Generator: 8-layer CNN Discriminator: 12-layer CNN | FDG simulated images for pre-training and human images for validation | Two L2 adversarial losses + cycle-consistency loss + total variation penalty | Input: low-resolution PET, T1-weighted MR imaging, spatial (radial + axial) coordinates Output: high-resolution PET Training target: unpaired high-resolution PET | PSNR, RMSE, SSIM |

Abbreviations: 2D: 2-dimensional, 4D: 4-dimensional, cGAN: conditional generative adversarial network, CNR: contrast-to-noise ratio, CRC: contrast recovery coefficient, FLAIR: fluid-attenuated inversion recovery, 18F-FMISO: 18F-fluoromisonidazole, GAN: generative adversarial network, MAPE: mean absolute percentage error, MLEM: maximum likelihood expectation maximization, MSE: mean square error, MSSIM: mean structural similarity index measure, NLM: nested logit model, NRMSE: normalized root-mean-square error, OSEM: ordered subsets expectation-maximization, PSNR: peak signal-to-noise ratio, RMSE: root-mean-square error, SSIM: structural similarity index.

## 2.2. Supervised Learning Models and Network Architectures

Almost all AI-based PET image processing techniques, including both supervised and unsupervised techniques feature some convolutional layers in the neural network. Convolutional neural networks (CNNs), which contain convolutional layers and could also contain other layer types (e.g., pooling or fully-connected layers), have emerged as a vital family of image processing tools in the AI era and have been leveraged extensively for PET image enhancement.

Neural networks for image processing are often characterized as 2-dimensional (2D), 2.5-dimensional (2.5D), or 3D based on the input dimensionality. 2D models receive as inputs 2D image slices from the whole image or 2D image patches (i.e., a 2D slice from a sub-image) in one or more orientations (i.e., axial, coronal, or sagittal). A 2.5D model uses image "slabs", i.e., sets of contiguous 2D slices, and carries some contextual information (i.e., information on the local neighborhood of a voxel) unlike their 2D counterpart. Fully 3D models receive 3D whole images or 3D image patches as inputs and tend to outperform 2D and 2.5D models [39, 45] but have steep memory requirements. Gong et al. proposed a mixed 2D and 3D encoder-decoder network to effectively synergize 2D and 3D features, which improved the fidelity of denoised images [43].

Many denoising approaches have exploited additional information from anatomical images. In many papers, noisy PET images and co-registered MR or CT images were provided to the network via distinct input channels [28, 57]. Anatomically guided approaches have been reported to improve image quality and quantitative accuracy. However, some works mentioned the potential performance degradation that could stem from inter-modality alignment errors [53].

Many of the first supervised learning techniques for PET image denoising were based on CNNs with shallower architectures, i.e., with relatively few layers. The CNN in da Costa-Luis et al., for example, was based on only 3 layers [27]. This network, termed µ-net, has relatively few unknown parameters and is well-adapted for settings with very limited training data. µ-net was demonstrated to be more robust than deeper networks which performed poorly in the validation phase due to overfitting of the training data [28]. Later CNN implementations for PET image denoising use deeper architectures, such as the implementation by Gong et al. [29]. Many CNN-based PET denoising architectures have benefitted from a residual learning strategy [66], which involves learning the difference between the clean and the corrupt images instead of directly learning the clean image [29]. Dilated convolutions have also been shown to improve PET image denoising performance [32, 67].

While most early image denoising efforts were based on pure CNNs, many recent architectural innovations have greatly boosted image denoising performance. A popular architecture widely used for many different image processing tasks today is the U-Net, which was originally proposed for image segmentation [68]. The U-Net architecture consists of a contracting path (resembling an encoder) which reduces the input dimensionality followed by an expansive path (resembling a decoder) with multiple skip connections between the paths that enable feature concatenation to aid upsampling. U-Net applications include supervised denoising models [34, 35, 42, 45] as well as unsupervised models [57-59, 61] to be discussed in section 2.3. A thorough comparison of different U-Net formats – 2D, 2.5D, and 3D – was conducted in a paper by Lu et al., which showed that a fully 3D U-Net leads to lower SUV bias than 2D or 2.5D alternatives [45]. A detailed discussion of neural network architectures can be found in Fereshteh Yousefirizi and colleagues' article, "Towards High-Throughput AI-Based Segmentation in Oncological PET Imaging," of this Special Issue.

Generative adversarial networks (GANs) [69] have rapidly gained popularity in the PET image reconstruction and processing arenas. A GAN consists of two competing neural networks, a generator and a discriminator, that are simultaneously trained. For an image enhancement task, the generator seeks to produce synthetic versions of clean images while the discriminator tries to distinguish between real and synthetic versions of the clean images. Upon completion of the training, the generator would be able to synthesize a highly realistic clean image from a corrupt image. While many PET image enhancement efforts based on GANs are unsupervised, GANs have been successfully used in supervised settings to synthesize realistic clean PET images from their noisy counterparts and have been shown to outperform CNN-based alternatives [41, 42, 54, 64, 65]. Wang et al. published a PET denoising technique based on 3D conditional GANs that uses an iterative refinement scheme for robust training [39]. Ouyang et al. proposed a GAN to denoise ultra-low-dose (1% of the standard dose) amyloid PET images and used a U-Net for the generator, a feature matching adversarial loss to reduce artifacts in the synthetic images, and a task-specific perceptual loss to boost the presence of accurate neuropathological details in the images [42]. In a paper on PET image denoising by Xue et al., residual learning and self-attention were reported to improve GAN performance by preserving structural details and edges in the images [38]. Gong et al. used Wasserstein distances to boost GAN performance at the PET image denoising task and relied on task-specific initialization based on transfer learning to reduce the training time [43]. Zhou et al. used a Siamese adversarial network to recover high-dose respiratory gated PET image volumes from their low-dose gated counterparts [70].

The CycleGAN [71], a specialized configuration of two pairs of GANs originally proposed for domain translation, has found widespread use in image synthesis and processing. CycleGANs for PET image enhancement consist of two generator-discriminator pairs, with the first generator performing the mapping from the corrupt image domain to the clean image domain and the second generator performing the reverse mapping. A cycle consistency loss is introduced to ensure parity between the synthetic and the original corrupt images. Zhou et al. proposed a CycleGAN for denoising low-dose oncological images that relies on the Wasserstein distance metric for the generator loss function in order to stabilize training and showed that this method leads to low bias in both the tumor and background tissues [41]. Fig. 3 shows the network architecture and sample denoised images from Zhou et al. [41]. Zhao et al. also use the CycleGAN for low-dose PET image restoration [37]. Sanaat et al. compared CycleGAN with a residual network for 1/8 low-dose whole-body PET image denoising [54]. The CycleGAN achieved better image quality than a residual network and led to similar performance in lesion detectability compared to full-dose images.

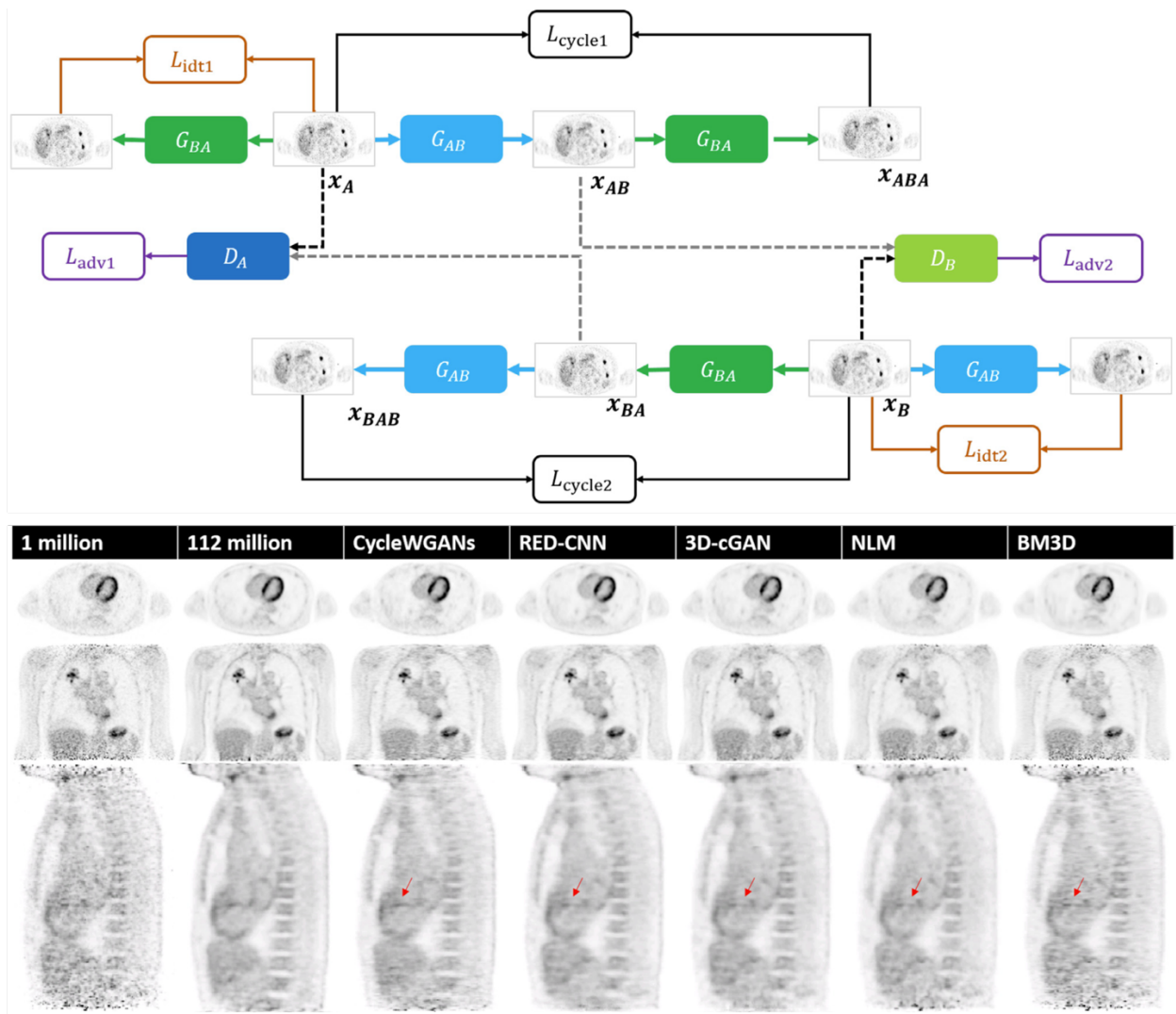

**Fig. 3.** Network architecture of a Wasserstein CycleGAN (referred to here as CycleWGAN) for supervised image denoising (top) and sample denoising results (bottom). (Adapted from Zhou, L., et al., Supervised learning with CycleGAN for low-dose FDG PET image denoising. Medical Image Analysis, 2020. 65: p. 101770. [41])

Similar to the denoising problem, solutions to the deblurring and super-resolution problems have been developed in the image processing and computer vision communities that rely on CNNs or more complex architectures like the U-Net, GAN, or CycleGAN [72-75]. Specifically, for PET image enhancement, Song et al. used a very deep CNN architectures with 20 layers [62]. Alongside low-resolution PET, the network received high-resolution anatomical MRI inputs to facilitate resolution recovery. High-resolution PET image patches were used as targets for network training. To accommodate the spatially varying nature of PET image resolution, radial and axial locations for each input patch were passed as additional inputs to the CNN. Training was performed in supervised mode for both simulation and clinical data, and it was shown that the very deep CNN with anatomical guidance outperforms penalized deconvolution as well as shallower CNNs.

## 2.3. Unsupervised, Weakly-Supervised, and Self-Supervised Learning Models and Network Architectures

The techniques discussed in the previous section are all based on supervised learning and require clean target images corresponding to each corrupt image for network training. In clinical settings, such paired datasets are difficult to come by. For denoising, this would require high-count images corresponding to each low-count image. Such image pairs can be generated by performing rebinning on high-count list-mode data [32, 76], but this step would require access to raw list-mode data. For the deblurring task, a supervised learning model would require a high-resolution image corresponding to each low-resolution image. This can be achieved either by performing two scans of the same subject, one on a state-of-the-art high-resolution scanner and another on a lower-resolution scanner, or by artificially degrading a high-resolution PET image using the measured or estimated PSF corresponding to the low-resolution input. These stringent data requirements limit the practical utility of supervised learning models. This has led to a steadily growing interest in unsupervised models for PET denoising and deblurring. It should be noted that the term *self-supervised* learning is often used to refer to some unsupervised techniques that learn the corrupt-to-clean mapping directly from the input image (e.g., from neighborhood information or population-level characteristics) without requiring a separate paired clean image. *Weakly supervised* methods, on the other hand, rely on noisy or imprecise targets for supervision.

The Noise2Noise technique is a promising weakly supervised denoising approach that reconstructs a clean image from multiple independent corrupt observations [77]. This method was applied to multiple noise realizations of a PET image with similar counts by Chan et al. to create a denoised PET image and was shown to be able to suppress noise while promoting a natural noise texture and reducing speckle and clustered noise [78]. Wu et al. applied Noise2Noise for denoising $^{15}$O-water dynamic PET images with a short half-life of only ~2 min, which tend to be extremely noisy, by using noisy time-frames from separate injections as the independent noise realizations required for the Noise2Noise method [60].

The Noise2Noise technique requires at least two noise realizations, which are not available for most clinical datasets, except for tracers with short half-lives that enables repeated scans with low total patient dose, such as $^{15}$O-water and $^{82}$Rb. In comparison, the Noisier2Noise approach trains the network with noisy and noisier image pairs, where the noisier image is derived from a single noise realization with added synthetic noise based on a known statistical model [79]. Yie et al. compared PET image denoising performance of Noisier2Noise with Noise2Noise and a supervised approach [61]. Their Noisier2Noise model was trained using PET images acquired over 10-s time frames as the input and corresponding 40-s duration images as the target. It was compared with Noise2Clean (input: 40-s duration images, training target: 300-s duration images) and Noise2Noise (pairs of 40-s duration images used for network training). Their results show that, while Noise2Noise and Noise2Clean demonstrate comparable denoising performance, Noisier2Noise is effective at noise suppression while maintaining the noise texture in the input.

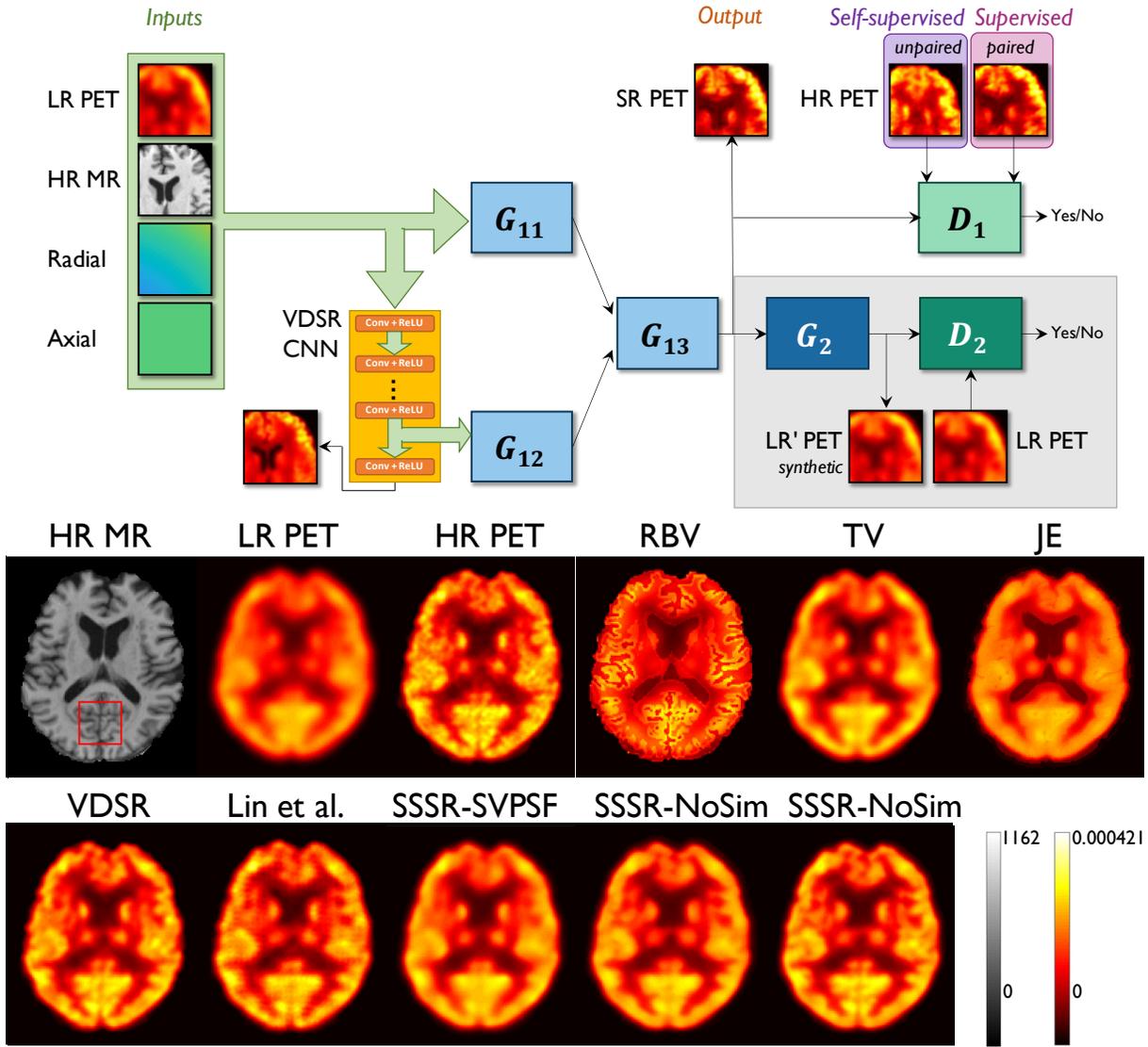

**Fig. 4.** Network architecture based on dual GANs for self-supervised image super-resolution (SSSR) (top) and sample deblurring results (bottom). Abbreviations used: HR - high-resolution, LR - low-resolution, JE - joint entropy, RBV - region-based voxel-wise, TV - total variation, VDSR - very deep super-resolution. (Adapted from Song, T.A., et al., PET image super-resolution using generative adversarial networks. Neural Networks, 2020. 125: p. 83-91. [65])

The deep image prior is another very popular and versatile technique which is unsupervised and was originally demonstrated to be useful for a wide range of image restoration problems [80]. This method is founded on the idea that the intrinsic structure of a generator network is able to capture low-level image statistics and can be used to regularize an image without any learning. Cui et al. used the deep image prior with a U-Net based generator to denoise PET images in an unsupervised manner. Instead of providing a random input to the generator, they provided an anatomical image at the input and reported notable improvements in the contrast-to-noise ratio [57]. Hashimoto et al. use the deep image prior for dynamic PET imaging denoising and used a less noisy static PET image as the input to a U-Net generator [59].

Unsupervised methods for PET image deblurring and partial volume correction are still emerging. Song et al. reported a dual GAN architecture somewhat similar to the CycleGAN for self-supervised resolution recovery of PET images using unpaired low- and high-resolution image sets [65]. Fig. 4 shows the network architecture and sample denoised images from Song et al. [65]. The network received as inputs a low-resolution PET image, a high-resolution anatomical MR image, spatial information (axial and radial coordinates), and a high-dimensional feature set extracted from an auxiliary CNN which was separately trained in a supervised manner using paired simulation datasets.

## 2.4. Evaluation Metrics

Traditionally, PET image reconstruction and enhancement techniques are assessed by studying the tradeoff between two statistical measures – bias and variance. The normalized root-mean-square error is an error metric with broad application in many domains and is frequently used in the context of PET as well. In addition, the mean absolute percentage error, peak signal-to-noise ratio and structural similarity index, two image quality metrics popular in the computer vision community, are frequently used nowadays to evaluate AI-based PET image enhancement techniques. In addition, metrics like the contrast recovery coefficient, contrast-to-noise ratio, and SUV bias are also widely used. In the following, the ground truth and estimated images are denoted $x$ and $\hat{x}$ respectively. We use the notations $\mu_x$ and $\sigma_x$ respectively for the mean and standard deviation of $x$ based on averaging over the voxels, with $N$ being the number of voxels in the whole image or over one ROI.

1. **Bias and Variance**: Many image enhancement techniques reduce noise in the images by adding bias. It is desirable to simultaneously achieve both low bias and low variance. When $M$ noise realizations of the enhanced image are available, the bias and variance at the $i$th voxel can be computed as follows:

$$\text{bias}_i = \frac{1}{M} \sum_{k=1}^{M} (\hat{x}_i^k - x_i),$$

$$\text{var}_i = \frac{1}{M} \sum_{k=1}^{M} \left( \hat{x}_i^k - \frac{1}{M} \sum_{j=1}^{M} (\hat{x}_i^j) \right)^2,$$

where $j$ and $k$ are sample indices. To study the bias-variance tradeoff, the bias and variance (for all voxels or within an ROI) can be plotted against each other. Among a set of image reconstruction or enhancement techniques being compared, the method with the lowest lying bias-variance curve is preferred. This is a very robust technique but requires the knowledge of the ground truth and multiple noise realizations, both of which are feasible in simulations.

2. **Normalized Root-Mean-Square Error (NRMSE)**: The NRMSE can be computed by normalizing the root-mean-square error by the mean:

$$\text{NRMSE}(\hat{x}, x) = \frac{\sqrt{\frac{1}{N} \sum_{i=1}^{N} (\hat{x}_i - x_i)^2}}{\mu_x}.$$

3. **Mean Absolute Percentage Error (MAPE)**: The MAPE can be computed as:

$$\text{MAPE}(\hat{x}, x) = \frac{1}{N} \sum_{i=1}^{N} \left| \frac{\hat{x}_i - x_i}{x_i} \right|.$$

4. **Peak Signal-to-Noise Ratio (PSNR)**: PSNR is defined as the ratio of the maximum signal, typically the maximum image voxel value that could be sensitive to noise, to the noise power:

$$\text{PSNR}(\hat{x}, x) = 10 \log_{10} \left( \frac{\max(x)^2}{\frac{1}{N} \sum_{i=1}^{N} (\hat{x}_i - x_i)^2} \right).$$

5. **Structural Similarity Index (SSIM)**: SSIM [81] is a well-accepted measure of the perceived consistency of the image and is computed as:

$$\text{SSIM}(\hat{x}, x) = \frac{(2\mu_{\hat{x}} \mu_x + C_1)(2\sigma_{\hat{x}x} + C_2)}{(\mu_{\hat{x}}^2 + \mu_x^2 + C_1)(\sigma_{\hat{x}}^2 + \sigma_x^2 + C_2)},$$

where $C_1$ and $C_2$ are parameters that are used to stabilize the division operation and $\sigma_{\hat{x}x}$ is the covariance of the $x$ and $\hat{x}$.

6. **Contrast Recovery Coefficient (CRC)**: The CRC for a target ROI $\mathcal{R}$ with respect to a reference region $\mathcal{R}_{\text{ref}}$ (e.g., a tumor vs. its background) is defined as follows:

$$\text{CRC} = \frac{\mu_x^{\mathcal{R}} - \mu_x^{\mathcal{R}_{\text{ref}}}}{\mu_x^{\mathcal{R}_{\text{ref}}}}.$$

7. **Contrast-to-Noise Ratio (CNR)**: The CNR for a target ROI $\mathcal{R}$ with respect to a reference region $\mathcal{R}_{\text{ref}}$ (e.g., a tumor vs. its background) is defined as follows:

$$\text{CNR} = \frac{\mu_x^{\mathcal{R}} - \mu_x^{\mathcal{R}_{\text{ref}}}}{\sqrt{\left[\left(\sigma_x^{\mathcal{R}}\right)^2 + \left(\sigma_x^{\mathcal{R}_{\text{ref}}}\right)^2\right]}}$$

To assess how well an enhanced image performs relative to the original corrupt image, some papers compute a CNR improvement ratio by taking the difference between the CNRs of the enhanced and original images and dividing it by the CNR of the original image.

8. **SUV Bias:** The SUV is a semi-quantitative metric that is widely used in clinical practice. The mean or the maximum SUV in an ROI $\mathcal{R}$ are both commonly used for ROI quantitation. To assess how well an enhanced image performs relative to the original corrupt image, the SUV change can be computed by taking the difference between the mean or maximum SUV of the enhanced and original images and dividing it by the corresponding number of the original image.

The above metrics are all task-independent metrics that assess image quality globally or for an ROI. PET images are often acquired for a specific clinical task, e.g., detection of a cancerous lesion. A general improvement in image quality may not always proportionately impact the accuracy of the final clinical task. This has led to the development of task-based image quality metrics that are application-dependent. Liu et al., for example, investigated the impact of noise reduction achieved via deep learning on lesion detectability as a function of count statistics based on a cross-center phantom study [82]. Their U-Net based denoiser substantially improved the detectability of small lesions in low-count PET images even though there was reduced contrast in both liver and lung lesions. In a study conducted by Schaefferkoetter et al., three experienced physicians ranked denoised PET images based on both subjective image quality and lesion detectability for different noise/count levels [34]. Their U-Net based denoiser outperformed Gaussian filtering in terms of physician-assigned image quality ranks with the largest improvement reported for the lowest-count images. When evaluated for lesion detectability – a task-specific metric, AI seemed to only be beneficial for the low-count images. For standard-count images, U-Net and conventional Gaussian filtered images led to similar lesion detectability outcomes. Xu et al. used an FDA-cleared AI-based proprietary software package SubtlePET™ to enhance low-dose noisy PET images of lymphoma patients and reported that standard-count and AI-enhanced low-count PET images led to similar lymphoma staging outcomes based on assessments by two physicians [83]. Katsari et al. also used SubtlePET™ to process $^{18}$F-FDG PET/CT examinations obtained with 66% standard dose [56]. Their results showed that there was no significant difference between AI-enhanced low-dose images and standard-dose images in terms of lesion detectability and SUV-based lesion quantitation. Nai et al. conducted a lesion-detection study using a PET/CT protocol for lung screening based on 30% of the standard radiotracer dose [84]. The study found that the AI-processed low-dose images achieved substantial image quality improvement and were able to preserve lesion detectability. Chen et al also found that AI-enhanced 1%-dose PET images achieved comparable accuracy for amyloid status determination as standard-dose images [55]. Tsuchiya et al. conducted a study involving $^{18}$F-FDG PET datasets from 50 patients and reported AI-based image enhancement significantly improved tumor delineation and overall image quality and suppressed image noise compared to a conventional Gaussian filter [52]. A more detailed description of task-based assessment of AI can be found in Abhinav K. Jha and colleagues' article, "Objective Task-Based Evaluation of Artificial Intelligence-Based Medical Imaging Methods: Framework, Strategies and Role of the Physician," of this Special Issue.

## 2.5. Loss Functions

The loss function plays an important role in network training. Table 1 specifies the loss function(s) used in each listed paper. The most common loss function used for image enhancement applications is the mean squared error loss, which is based on the L2 norm of the difference between the network output and the training target and is also referred to as the *L2 loss*. A common alternative is the *L1 loss*, which is based on the L1 norm and is less sensitive to outliers than the L2 loss. Some papers have used an L2 or L1 loss term based on the difference between the gradients of the output and target as a way to better preserve structural information in images [40, 48]. A number of loss functions compare probability distributions instead of individual datapoints. One such common measure from information theory widely used as a loss function in machine learning is the *Kullback-Leibler divergence*, which measures how one probability distribution diverges from a second expected probability distribution. Another common loss function, the *Jensen-Shannon divergence*, is based on the Kullback-Leibler divergence but is a symmetric measure with an upper bound of 1. This divergence metric computed between the generator output and observed data distributions is commonly minimized for GAN training. The *Wasserstein distance*, also known as the earth mover's distance, has gained popularity as a loss function for many image enhancement problems. Several of the GAN-based denoising papers discussed in this article and highlighted in Table 1 rely on a Wasserstein loss function. The commonly used evaluation metric SSIM and its variant multi-scale SSIM have also been used as a loss function in certain applications. Another notable loss function for many imaging applications, including PET, is the perceptual loss [85] which is based on differences between high-level image feature representations extracted by a given network those from a pre-trained CNN (mostly commonly a pre-trained VGG16 network [86]. Perceptual loss has been shown to be very effective at improving the visual quality of images.

## 3. Emerging Approaches and Novel Applications

### 3.1. Overview

The past trends in AI-based PET image enhancement have been presaged by related developments in the broader machine learning community. Some areas of emphasis in image processing and computer vision research today include the development of approaches that enable the collation and collective use of datasets across multiple centers to ensure higher volume and variety of training data and improved generalization of the resultant models. Another focal area in the machine learning field is the development of models that are interpretable. These efforts are only beginning to percolate through the medical imaging arena with relatively few applications demonstrated on PET imaging at this point. In the following sub-sections, we will discuss some of these areas which are either beginning to impact PET imaging or are likely to do so in the near future.

### 3.2. Transfer Learning

Transfer learning involves training a model on one (usually larger) dataset and fine-tuning it for a related application using a different dataset [87, 88]. While transfer learning is not a new approach, it continues to find novel applications in the medical imaging field. It is particularly valuable in applications where the training data is limited, a scenario very commonly encountered in clinical settings. The image denoising technique by Gong et al. described earlier, for example, uses a model pre-trained with simulation data and subsequently fine-tuned with real data [29]. As can be seen from Table 1, the overwhelming majority of AI-based efforts in PET are based on the $^{18}$F-FDG radiotracer. Smaller data sizes for non-FDG radiotracers and/or extremely long/short half-lives of some radioisotopes are impediments to the training of AI models based on radiotracers other than $^{18}$F-FDG. Liu et al. used transfer learning to extend denoising models trained on $^{18}$F-FDG PET images to data based on two less-used tracers: $^{18}$F-fluoromisonidazole (commonly referred to as FMISO) and $^{68}$Ga-DOTATATE [44]. They reported that fine-tuning greatly reduced NRMSE and ROI bias and improved the signal-to-noise ratio compared to cases with no transfer learning. Chen et al. used transfer learning to generalize a deep learning model to overcome data bias in multi-site PET/MR studies where there are disparities in acquisition protocols and reconstruction parameters [89]. In this study, a U-Net network was pretrained using data from a GE SIGNA scanner with time-of-flight capability and subsequently finetuned using the data from a Siemens mMR scanner.

## 3.3. Federated Learning

Data security and privacy issues are of great importance in the medical domain and pose a formidable barrier to the collation of large multi-center patient databases for the effective training and validation of AI models. Usually, in machine learning based on multi-site data, the data from each location are uploaded to a centralized server and subsequently used for model training and validation. Federated learning is a novel and increasingly popular alternative to the traditional centralized learning paradigm and involves distributing the model training task across sites and eventually aggregating the results [90]. It allows collaborative and decentralized training of deep learning models without any patient data sharing. In federated learning, each site uses its local data to train the model and shares the trained model parameters with a centralized server which builds an aggregate model that is shared back with the individual sites. Federated learning has become a common trend in the medical imaging domain by enabling access to distributed data for training deep learning models while abiding by patient privacy regulations. Applications of federated learning in medical imaging are gradually emerging. Its utility has been demonstrated in the context of classification problems based on MRI and CT imaging. For example, Li et al. used federated learning for MRI-based brain tumor segmentation [91]. Additionally, Kumar et al. and Dou et al. used federated learning models for COVID-19 detection using chest CT scans [92, 93]. While federated learning has not been extensively investigated for PET, the multi-site transfer learning study by Chen et al. involved model sharing across sites and compared finetuned models with directly trained models for site-specific datasets [89].

## 3.4. Data Harmonization

Data heterogeneity is a major bottleneck while processing multi-center medical imaging studies. Differences in image acquisition protocols (including scan time), scanner models, image reconstruction parameters, and post-reconstruction processing could lead to substantial variability in image characteristics, e.g., image size, field-of-view, spatial resolution, and noise. Scan time differences in acquisition protocols could lead to drastically different noise levels across sites. Collective processing of such image pools often involves aggressive Gaussian filtering that would render the lowest-count images usable but over-smoothing the higher-count images in the process. In many longitudinal multi-site PET studies, some sites stick to lower-resolution legacy scanners for temporal consistency while others have access to state-of-the-art high-resolution scanners. In such scenarios, a common harmonization strategy involves degrading higher-resolution images to match the spatial resolution of the lowest-resolution images. The PET image denoising and deblurring techniques discussed in this review all have broad applicability for image harmonization as they enable improving the resolution and noise characteristics of the lowest-quality images instead of degrading the highest-quality images. The super-resolution techniques published by Song et al. [62, 65] are useful for data harmonization as they focus on developing a mapping from a low-resolution image domain (of a legacy scanner) to a high-resolution image domain (of a state-of-the-art dedicated brain PET scanner).

Besides harmonization between images, PET is unique in that there is potential for using AI for "within-image" harmonization. Notably, the resolution of PET images is spatially variant with the resolution at the periphery typically much worse than that at the center of the image. Such resolution nonuniformities pose a severe challenge to PET quantitation. Over the years, the PET community has taken great interest in spatially-variant regularization techniques that generate uniform resolution PET images [94, 95]. AI-based PET image enhancement techniques that achieve uniform resolution are yet to be explored, but these methods may offer an accessible, purely post-processing-based solution to a longstanding problem in PET.

### 3.5. Reinforcement Learning

There are three fundamental machine learning paradigms: supervised, unsupervised learning, and reinforcement learning. While the first two have been discussed here at length, reinforcement learning is only beginning to be applied for image enhancement and restoration. In reinforcement learning, an artificial agent interacts with its environment through a series of actions intended to maximize the reward for a given task. Smith et al. reported using reinforcement learning for PET image segmentation [96], where the main objective was to detect the location of a lesion by moving a bounding box in different directions in the image to maximize a cumulative reward. The reward function in this scenario was a measure of the agreement between the predicted location of the detected lesion that of the one predefined by radiologists. While reinforcement learning has been used for image segmentation, its applications to image enhancement are gradually emerging. The pixelRL technique is one such approach, which was shown to work for a variety of image processing problems [97]. This method poses an image enhancement tasks as a multi-agent reinforcement learning problem, where each pixel has an agent which can alter the pixel intensity by taking an action intended to maximize a reward. Other examples of reinforcement learning used for image enhancement include a superkernel approach proposed for image denoising [98] and the RL-Restore technique for image restoration [99]. While medical image enhancement based on reinforcement learning is in a nascent stage, it is a most promising research direction for PET. PET image enhancement, for example, could be formulated as a multi-agent reinforcement learning problem where the intensity (i.e., the current state) of each voxel can be iteratively updated by means of an agent's action. The rewards can be defined as either image-based metrics or clinical task-based metrics. By defining rewards that are based on clinical tasks, it may be possible to integrate clinical decision making with image enhancement, thereby making it easier to assess the clinical benefit of the image enhancement scheme.

### 3.6. Interpretability

The "black-box" format of AI models sometimes makes them less attractive in a clinical setting where intuitive explanations are sought by the clinical end-users. Emerging interpretable or explainable AI techniques, therefore, have garnered high levels of interest in the clinic. Most interpretable AI tools for medical imaging have focused on image classification problems [100-104]. Current computer vision literature shows a rising trend in papers on interpretable AI models for image deblurring, particularly blind image deblurring. In this category, unrolled deep networks are notable. Unrolled deep networks incorporate a physical model based on with prior domain knowledge, which makes them fundamentally interpretable. They typically have efficient implementations free of the usual computational costs and inaccuracies associated with model inversion [105-107]. Although the unrolling methods explored so far in PET are in the context of image reconstruction, unrolled networks are equally promising for post-reconstruction PET image enhancement techniques like denoising and deblurring as they can incorporate knowledge of Poisson statistics (for image denoising) and measured spatially variant point spread functions (for image deblurring). Attention-based approaches have been widely used to create explainable AI models, and recently they have been applied to image super-resolution [108]. Manifold modeling in embedded space has been proposed as an interpretable alternative to the deep image prior with broad applications in image denoising, super-resolution, and other inverse problems [109]. These methods are both highly relevant and easily extensible to PET imaging and, therefore, are likely to assume a more prominent role in PET image enhancement in the near future.

## 4. Conclusion

The excitement around AI in the PET imaging field is no surprise given the promising early results for the ill-posed inverse problems that have plagued the field since its infancy. Most papers discussed here demonstrate that AI-based approaches for PET image enhancement outperform conventional methods by a striking margin. Despite the rich variety in AI efforts for PET enhancement, a key challenge that we are facing today stems from the poor generalizability of most trained models. Data size and variability remain challenges to model training using medical image data and often lead to models that do not generalize well to diverse datasets. A performance drop is often observed when a trained network is applied to data from a different domain (i.e., different scanner, different scanning protocol, different tracer, different reconstruction settings). For example, in PET denoising, when the AI model is trained on PET images with a specific noise level, it could have insufficient denoising in noisier PET images or over-denoising/oversmoothing in less noisy PET images. Specifically, in the context of PET, there is interest in models that perform well on images with resolution and noise characteristics that are very different from the training data. While the recently proposed unsupervised denoising and deblurring models are significant steps to circumventing this challenge, there remains a performance gap between supervised and unsupervised approaches that needs to be bridged. Federated learning also offers the promise to surmount formidable barriers to data access and could lead to the development of more robust models, the performance of which can be replicated in data from multiple sites. Developments in this direction would be vital for engendering confidence in AI-based models and in ensuring the long-term success of AI in the PET imaging field.

While the review focus on AI-based post-reconstruction PET image enhancement, advanced sinogram denoising and PET image reconstruction techniques are vital to improve the image quality for post-reconstruction image enhancement. Sanaat et al. found that AI-based sinogram denoising outperformed AI-based image space denoising with higher image quality and lower SUV bias and variance [47]. The integration of AI-based reconstruction and post-reconstruction image enhancement has the potential to push advances in the performance of the underlying technology. Developments of intelligent image acquisition, reconstruction, analysis, detection, and diagnosis would release the super-power of AI.

While AI models have shown high performance when adjudged solely based on image quality measures, the performance margins over traditional approaches get slimmer for task-based measures. Unlike the models discussed in this paper, which receive corrupt images as inputs and create enhanced images at outputs, there also exist many "end-to-end" approaches that start with raw data, extract features, and directly output a clinical decision. While this latter group of models are optimized for clinical tasks, these are mostly black-box approaches that do not generate enhanced images. Interpretable image enhancement models that adopt an intermediate path which merges the image enhancement task with one or more clinical decision-making tasks could bridge the existing gap in knowledge and lead to goal-oriented image enhancement. Interpretability would also be vital for ensuring the wider adoptability of AI-based models in the clinic as it would address skepticism among clinicians about the black-box nature of these models.

The community would also benefit from standardized datasets for benchmarking AI models for image enhancement. This would address the disjointedness of current efforts and allow approaches to be evaluated against a common standard. While code sharing is an increasingly common practice in the PET community nowadays, the sharing of trained models would also be vital for fair evaluation of emerging approaches. As in most other research areas, AI-based methods in the PET field today have appreciable momentum. There is reason to be hopeful that creative solutions to the aforementioned problems await us in the near future. With more diverse datasets, rigorous and reproducible validation standards, and innovative models that integrate with existing clinical pipelines, AI-based models for PET image enhancement could ultimately lead to improved diagnosis, prognosis, and treatment evaluation and increasingly personalized approaches to healthcare.

## Acknowledgment
Juan Liu, Niloufar Mirian, and Chi Liu are supported by NIH grant R01EB025468. Masoud Malekzadeh, Tzu-An Song, and Joyita Dutta are supported in part by NIH grants K01AG050711, R21AG068890, and R03AG070750.

## References

1. Bar-Shalom, R., A.Y. Valdivia, and M.D. Blaufox, *PET imaging in oncology.* Semin Nucl Med, 2000. 30(3): p. 150-85.
2. Politis, M. and P. Piccini, *Positron emission tomography imaging in neurological disorders.* J Neurol, 2012. 259(9): p. 1769-80.

3. Knaapen, P, De Haan, S, Hoekstra, OS, et al. *Cardiac PET-CT: advanced hybrid imaging for the detection of coronary artery disease.* Neth Heart J 2010; 18(2): 90-8.
4. Shepp, L.A. and Y. Vardi, *Maximum Likelihood Reconstruction for Emission Tomography.* IEEE Transactions on Medical Imaging, 1982. 1(2): p. 113-122.
5. Dutta, J., S. Ahn, and Q. Li, *Quantitative statistical methods for image quality assessment.* Theranostics, 2013. 3(10): p. 741-756.
6. Leahy, R.M. and J. Qi, *Statistical approaches in quantitative positron emission tomography.* Statistics and Computing, 2000. 10(2): p. 147-165.
7. Panin VY, Kehren F, Michel C, Casey M. *Fully 3-D PET reconstruction with system matrix derived from point source measurements.* IEEE Trans on Med Imaging 2006; 25(7):907-21.
8. Vargas, P.A. and P.J. La Rivière, *Comparison of sinogram- and image-domain penalized-likelihood image reconstruction estimators.* Medical physics, 2011. 38(8): p. 4811-4823.
9. Tong, S., A.M. Alessio, and P.E. Kinahan, *Image reconstruction for PET/CT scanners: past achievements and future challenges.* Imaging Med, 2010. 2(5): p. 529-545.
10. Tomasi C, Manduchi R. *Bilateral filtering for gray and color images. Sixth International conference on computer vision* (IEEE cat. No.98CH36271), Bombay, India. Jan 7, 1998. pp.839–46.
11. Hofheinz F, Langner J, Beuthien-Baumann B, et al. *Suitability of bilateral filtering for edge-preserving noise reduction in PET.* EJNMMI Res 2011; 1(1):23.
12. Perona, P. and J. Malik, *Scale-space and edge detection using anisotropic diffusion.* IEEE Transactions on Pattern Analysis and Machine Intelligence, 1990. 12(7): p. 629-639.
13. Antoine MJ, Travere JM, Bloyet D. *Anisotropic diffusion filtering applied to individual PET activation images: a simulation study.* In 1995 IEEE Nuclear Science Symposium and Medical Imaging Conference (NSS/MIC). San Francisco, United States. Oct 21-28, 1995. Vol. 3, pp.1465-1469.
14. Stefan W, Chen K, Guo H, et al. *Wavelet-based de-noising of positron emission tomography scans.* J Sci Comput 2012;50(3):665–77..
15. Turkheimer FE, Aston JA, Banati RB, et al. *A linear wavelet filter for parametric imaging with dynamic PET.* IEEE Trans Med Imaging 2003; 22(3):289–301.
16. Su, Y. and K.I. Shoghi, *Wavelet denoising in voxel-based parametric estimation of small animal PET images: a systematic evaluation of spatial constraints and noise reduction algorithms.* Phys Med Biol, 2008. 53(21): p. 5899-915.

17. Le Pogam A, Hanzouli H, Hatt M, et al. *Denoising of PET images by combining wavelets and curvelets for improved preservation of resolution and quantitation.* Med Image Anal 2013;17(8):877–91.
18. Arabi, H. and H. Zaidi, *Improvement of image quality in PET using post-reconstruction hybrid spatial-frequency domain filtering.* Phys Med Biol, 2018. 63(21): p. 215010.
19. Buades A, Coll B, Morel J. *A non-local algorithm for image denoising. In 2005 IEEE Computer Soc Conf Computer Vis Pattern Recognition (CVPR).* San Diego, United States. June 20-26, 2005. Vol. 2, pp. 60-65.
20. Dutta, J., R.M. Leahy, and Q. Li, *Non-local means denoising of dynamic PET images.* PloS one, 2013. 8(12): p. e81390-e81390.
21. Chan C, Fulton R, Barnett R, et al. *Postreconstruction nonlocal means filtering of whole-body PET with an anatomical prior.* IEEE Trans Med Imaging 2014;33(3):636–50.
22. Kirov, A.S., J.Z. Piao, and C.R. Schmidtlein, *Partial volume effect correction in PET using regularized iterative deconvolution with variance control based on local topology.* Phys Med Biol, 2008. 53(10): p. 2577-91.
23. Song TA, Yang F, Chowdhury SR, et al. *PET image Deblurring and superresolution with an MR-based joint entropy prior.* IEEE Trans Comput Imaging 2019;5(4):530–9.
24. Rousset, O.G., Y. Ma, and A.C. Evans, *Correction for partial volume effects in PET: principle and validation.* J Nucl Med, 1998. 39(5): p. 904-11.
25. Thomas BA, Erlandsson K, Modat M, et al. *The importance of appropriate partial volume correction for PET quantification in Alzheimer's disease.* Eur J Nucl Med Mol Imaging 2011;38(6):1104–19.
26. Müller-Gärtner HW, Links JM, Prince JL, et al. *Measurement of radiotracer concentration in brain gray matter using positron emission tomography: MRI-based correction for partial volume effects.* J Cereb Blood Flow Metab 1992;12(4):571–83.
27. Costa-Luis COD, Reader AJ. *Convolutional micronetworks for MR-guided low-count PET image processing.* In 2018 IEEE nuclear science symposium and medical imaging conference proceedings (NSS/MIC). Sydney, Australia. Nov 10-17, 2018. pp.1-4.
28. Costa-Luis, C.O.d. and A.J. Reader, *Micro-Networks for Robust MR-Guided Low Count PET Imaging.* IEEE Transactions on Radiation and Plasma Medical Sciences, 2021. 5(2): p. 202-212.
29. Gong K, Guan J, Liu CC, et al. *PET image denoising using a deep neural network through fine tuning.* IEEE Trans Radiat Plasma Med Sci 2019;3(2):153–61.

30. Xiang L, Qiao Y, Nie D, et al. *Deep auto-context convolutional neural networks for standard-dose PET image estimation from low-dose PET/MRI.* Neurocomputing 2017; 267:406–16.
31. Chen KT, Gong E, de Carvalho Macruz FB, et al. *Ultra-low-dose (18)F-florbetaben amyloid PET imaging using deep Learning with multi-contrast MRI inputs.* Radiology 2019; 290(3): 649–56.
32. Spuhler K, Serrano-Sosa M, Cattell R, et al. *Full-count PET recovery from lowcount image using a dilated convolutional neural network.* Med Phys 2020;47(10):4928–38.
33. Serrano-Sosa M, Spuhler K, DeLorenzo C, et al. *Denoising low-count PET images Using a dilated convolutional neural network for kinetic modeling.* J Nucl Med 2020; 61(supplement 1):437.
34. Schaefferkoetter J, Yan J, Ortega C, et al. *Convolutional neural networks for improving image quality with noisy PET data.* EJNMMI Res 2020;10(1):105.
35. Sano A, Nishio T, Masuda T, et al. *Denoising PET images for proton therapy using a residual U-net.* Biomed Phys Eng Express 2021; 7(2): 025014.
36. Wang Y, Zhou L, Yu B, et al. *3D auto-context-based locality adaptive multi-modality GANs for PET synthesis.* IEEE Trans Med Imaging 2019;38(6):1328–39.
37. Zhao K, Zhou L, Gao S, et al. *Study of low-dose PET image recovery using supervised learning with CycleGAN.* PLoS One 2020;15(9): e0238455.
38. Xue H, Teng Y, Tie C, et al. *A 3D attention residual encoder–decoder least-square GAN for low-count PET denoising.* Nucl Instrum Methods Phys Res 2020; 983:164638.
39. Wang Y, Yu B, Wang L, et al. *3D conditional generative adversarial networks for high-quality PET image estimation at low dose.* Neuroimage 2018; 174:550–62.
40. Kaplan, S. and Y.M. Zhu, *Full-Dose PET Image Estimation from Low-Dose PET Image Using Deep Learning: a Pilot Study.* J Digit Imaging, 2019. 32(5): p. 773-778.
41. Zhou L, Schaefferkoetter JD, Tham IW, et al. *Supervised learning with CycleGAN for low-dose FDG PET image denoising.* Med Image Anal 2020; 65:101770.
42. Ouyang J, Chen KT, Gong E, et al. *Ultra-low-dose PET reconstruction using generative adversarial network with feature matching and task-specific perceptual loss.* Med Phys 2019;46(8):3555–64.
43. Gong Y, Shan H, Teng Y, et al. *Parameter-transferred Wasserstein generative adversarial network (PT-WGAN) for low-dose PET image denoising.* IEEE Trans Radiat Plasma Med Sci 2021;5(2):213–23.
44. Liu H, Wu J, Lu W, et al. *Noise reduction with cross-tracer and cross-protocol deep transfer learning for low-dose PET.* Phys Med Biol 2020; 65(18):185006.
45. Lu W, Onofrey JA, Lu Y, et al. *An investigation of quantitative accuracy for deep learning based denoising in oncological PET.* Phys Med Biol 2019;64(16): 165019.

46. Ladefoged CN, Hasbak P, Hornnes C, et al. *Low-dose PET image noise reduction using deep learning: application to cardiac viability FDG imaging in patients with ischemic heart disease.* Phys Med Biol 2021; 66(5): 054003.
47. Sanaat A, Arabi H, Mainta I, et al. *Projection space implementation of deep learning-guided low-dose brain PET imaging improves performance over implementation in image space.* J Nucl Med 2020;61(9):1388–96.
48. He Y, Cao S, Zhang H, et al. *Dynamic PET image denoising with deep learning-based joint filtering.* IEEE Access 2021; 9:41998–2012.
49. Wang YJ, Baratto L, Hawk KE, et al. *Artificial intelligence enables whole-body positron emission tomography scans with minimal radiation exposure.* Eur J Nucl Med Mol Imaging 2021; 48(9): 2771–2781
50. Schramm G, Rigie D, Vahle T, et al. *Approximating anatomically guided PET reconstruction in image space using a convolutional neural network.* Neuroimage 2021; 224:117399.
51. Jeong YJ, Park HS, Jeong JE, et al. *Restoration of amyloid PET images obtained with short-time data using a generative adversarial networks framework.* Sci Rep 2021; 11(1): 4825.
52. Tsuchiya J, Yokoyama K, Yamagiwa K, et al. *Deep learning-based image quality improvement of (18)F-fluorodeoxyglucose positron emission tomography: a retrospective observational study.* EJNMMI Phys 2021; 8(1):31.
53. Liu, C.C. and J. Qi, *Higher SNR PET image prediction using a deep learning model and MRI image.* Phys Med Biol, 2019. 64(11): p. 115004.
54. Sanaat A, Shiri I, Arabi H, et al. *Deep learning-assisted ultra-fast/low-dose whole-body PET/CT imaging.* Eur J Nucl Med Mol Imaging 2021; 48(8): 2405–2415.
55. Chen KT, Toueg TN, Koran ME, et al. *True ultra-low-dose amyloid PET/MRI enhanced with deep learning for clinical interpretation.* Eur J Nucl Med Mol Imaging 2021; 48(8): 2416–2425.
56. Katsari K, Penna D, Arena V, et al. *Artificial intelligence for reduced dose 18F-FDG PET examinations: a real-world deployment through a standardized framework and business case assessment.* EJNMMI Phys 2021; 8(1): 25.
57. Cui J, Gong K, Guo N, et al. *PET image denoising using unsupervised deep learning.* Eur J Nucl Med Mol Imaging 2019; 46(13):2780–9.
58. Hashimoto F, Ohba H, Ote K, et al. *Dynamic PET image denoising using deep convolutional neural networks without prior training datasets.* IEEE Access 2019;7: 96594–603.
59. Hashimoto F, Ohba H, Ote K, et al. *4D deep image prior: dynamic PET image denoising using an unsupervised four-dimensional branch convolutional neural network.* Phys Med Biol 2021; 66(1):015006.

60. Wu D, Gong K, Kim K, et al. *Deep denoising of O-15 water dynamic PET images without training data.* J Nucl Med 2020;61(supplement 1):433.
61. Yie SY, Kang SK, Hwang D, et al. *Self-supervised PET denoising.* Nucl Med Mol Imaging 2020; 54(6): 299–304.
62. Song TA, Chowdhury SR, Yang F, et al. *Super-resolution PET imaging using convolutional neural networks.* IEEE Trans Comput Imaging 2020; 6: 518–28.
63. Garehdaghi F, Meshgini S, Afrouzian R, et al. *PET image super resolution using convolutional neural networks.* In 2019 5th Iranian Conference on Signal Processing and Intelligent Systems (ICSPIS). Shahroud, Iran. Dec 18-19, 2019. pp. 1-5.
64. Chen, W.-j. and A. McMillan, *Single Subject Deep Learning-based Partial Volume Correction for PET using Simulated Data and Cycle Consistent Networks.* Journal of Nuclear Medicine, 2020. 61(supplement 1): p. 520-520.
65. Song TA, Chowdhury SR, Yang F, et al. *PET image super-resolution using generative adversarial networks.* Neural Netw 2020; 125:83–91.
66. He K, Zhang X, Ren S, et al. *Deep residual learning for image recognition.* In 2016 IEEE conference on computer vision and pattern recognition (CVPR). Las Vegas, United States. Jun 27-30, 2016. pp. 770-778.
67. Serrano-Sosa M, Spuhler K, DeLorenzo C, et al. *PET image denoising using structural MRI with a novel dilated convolutional neural network.* J Nucl Med 2020; 61(supplement1):434.
68. Ronneberger O, Fischer P, Brox T. *U-net: convolutional networks for biomedical image segmentation.* In 2015 Medical image computing and computer assisted Intervention conference (MICCAI). Cham, Switzerland. Oct 5-9, 2015. pp. 234-241. Springer Cham
69. Goodfellow IJ, Pouget-Abadie J, Mirza M, et al. *Generative adversarial nets.* In: Proceedings of the 27th international Conference on neural information processing systems, vol. 2. Montreal, Canada: MIT Press; 2014. p.2672–80.
70. Zhou B, Tsai YJ, Liu C. *Simultaneous denoising and motion estimation for low-dose gated PET using a Siamese adversarial network with gate-to-gate consistency learning.* In 2020 International conference on medical image computing and computer assisted Intervention conference (MICCAI). Lima, Peru, Virtual/Online. Oct 4-8, 2020. pp. 743-752. Springer Cham.
71. Zhu J-Y, Park T, Isola P, et al. *Unpaired image-to-image translation using cycle-consistent adversarial networks.* In 2017 IEEE conference on computer vision and pattern recognition (CVPR). Hawaii, United States. Jul 21-26, 2017. pp. 2223-2232.
72. Dong C, Loy CC, He K, et al. *Image super-resolution using deep convolutional networks. IEEE transactions on pattern analysis and machine intelligence.* 2015; 38(2): 295-307.

73. Kim J, Lee JK, Lee KM. *Accurate image super-resolution using very deep convolutional networks.* In 2016 IEEE conference on computer vision and pattern recognition (CVPR). Las Vegas, United States. Jun 27-30, 2016. pp. 1646-1654.
74. Lim B, Son S, Kim H, et al. *Enhanced deep residual networks for single image super-resolution.* In 2017 IEEE conference on computer vision and pattern recognition (CVPR). Hawaii, United States. Jul 21-26, 2017. pp. 136-144.
75. Ledig C, Theis L, Huszár F, et al. *Photo-realistic single image super-resolution using a generative adversarial network.* In 2017 IEEE conference on computer vision and pattern recognition (CVPR). Hawaii, United States. Jul 21-26, 2017. pp. 4681-4690.
76. Kim K, Wu D, Gong K, et al. *Penalized PET reconstruction using deep learning prior and local linear fitting.* IEEE Trans Med Imaging 2018;37(6):1478–87.
77. Lehtinen J, Munkberg J, Hasselgren J, et al. *Noise2Noise: learning image restoration without clean data.* In Proceedings of the 35th International conference on machine learning (ICML). Stockholm, Sweden: PMLR; Jul 15-18, 2018. vol. 80. p. 2965–74.
78. Chan C, Zhou J, Yang L, et al. *Noise to noise ensemble learning for PET image denoising.* In 2019 IEEE Nuclear Science Symposium and Medical Imaging Conference (NSS/MIC). Manchester, United Kingdom. Oct 26-Nov 2, 2019. pp. 1-3.
79. Moran N, et al. *Noisier2Noise: learning to denoise from unpaired noisy data.* In 2020 IEEE/CVF conference on computer vision and pattern recognition workshops (CVPRW). Seattle, United States, Virtual/Online. Jun 19, 2020. pp. 12064-12072.
80. Ulyanov, D., A. Vedaldi, and V. Lempitsky, *Deep Image Prior.* International Journal of Computer Vision, 2020. 128(7): p. 1867-1888.
81. Zhou W, Bovik AC, Sheikh HR, et al. *Image quality assessment: from error visibility to structural similarity.* IEEE Trans Image Process 2004; 13(4): 600–12.
82. Liu H, Viswanath V, Karp J, et al. *Investigation of lesion detectability using deep learning based denoising methods in oncology PET: a cross-center phantom study.* J Nucl Med 2020;61(supplement 1):430.
83. Xu F, Pan B, Zhu X, et al. *Evaluation of deep learning-based PET image enhancement method in diagnosis of lymphoma.* J Nucl Med 2020; 61(supplement 1):431.
84. Nai YH, Schaefferkoetter J, Fakhry-Darian D, et al. *Validation of low-dose lung cancer PET-CT protocol and PET image improvement using machine learning.* Phys Med 2021; 81:285–94.
85. Johnson J, Alahi A, Fei-Fei L. *Perceptual losses for real-time style transfer and super-resolution.* In 2016 European conference on computer vision (ECCV). Amsterdam, Netherlands. Oct 8-16, 2016. pp. 694-711. Springer science.
86. Simonyan K, Zisserman A. *Very deep convolutional networks for large-scale image recognition.* arXiv preprint arXiv: 2014, 1409.1556.

87. Zhuang F, Qi Z, Duan K, et al. *A comprehensive survey on transfer learning.* Proc IEEE 2021; 109: 43–76.
88. Pan, S.J. and Q. Yang, *A Survey on Transfer Learning.* IEEE Trans. on Knowl. and Data Eng., 2010. 22(10): p. 1345–1359.
89. Chen KT, Schürer M, Ouyang J, et al. *Generalization of deep learning models for ultra-low-count amyloid PET/MRI using transfer learning.* Eur J Nucl Med Mol Imaging 2020; 47(13):2998–3007.
90. ShellerMJ, Edwards B, Reina GA, et al. *Federated learning in medicine: facilitating multi-institutional collaborations without sharing patient data.* Scientific Rep 2020; 10(1): 12598.
91. Li W, Milletarì F, Xu D, et al. *Privacy-preserving federated brain tumor segmentation.* In 2019 International workshop on machine learning in medical imaging (MLMI). Shenzhen, China, Oct 13, 2019. pp. 133-141. Springer, Cham.
92. Kumar R, Khan AA, Kumar J, et al. *Blockchain-Federated-Learning and deep learning Models for COVID-19 detection using CT imaging.* IEEE Sens J. 2021; 21(14): pp. 16301-16314.
93. Dou Q, So TY, Jiang M, et al. *Federated deep learning for detecting COVID-19 lung abnormalities in CT: a privacy preserving multinational validation study.* NPJ Digital Med 2021; 4(1):60.
94. Qi, J. and R.M. Leahy, *Resolution and noise properties of MAP reconstruction for fully 3-D PET.* IEEE Trans Med Imaging, 2000. 19(5): p. 493-506.
95. Stayman, J.W. and J.A. Fessler, *Regularization for uniform spatial resolution properties in penalized-likelihood image reconstruction.* IEEE Trans Med Imaging, 2000. 19(6): p. 601-15.
96. Smith RL, Ackerley IM, Wells K, et al. *Reinforcement learning for object detection in PET imaging.* In 2019 IEEE nuclear science symposium and medical imaging conference (NSS/MIC). Manchester, United Kingdom. Oct 26-Nov 2, 2019. pp. 1-4.
97. Furuta, R., N. Inoue, and T. Yamasaki, *PixelRL: Fully Convolutional Network With Reinforcement Learning for Image Processing.* IEEE Transactions on Multimedia, 2020. 22(7): p. 1704-1719.
98. Mozejko M, Latkowski T, Treszczotko L, et al. *Super-kernel neural architecture search for image denoising.* In 2020 IEEE/CVF conference on computer vision and pattern recognition workshops (CVPRW). Seattle, United States, Virtual/Online. Jun 19, 2020. pp. 484-485.
99. Yu K, Dong C, Lin L, et al. *Crafting a toolchain for image restoration by deep reinforcement learning.* In 2018 IEEE/CVF conference on computer vision and pattern recognition (CVPR). Salt Lake, United States, June 18-22, 2018. pp. 2443-2452.

100. Qiu S, Joshi PS, Miller MI, et al. *Development and validation of an interpretable deep learning framework for Alzheimer's disease classification.* Brain 2020; 143(6):1920–33.
101. Lee E, Choi JS, Kim M, et al. *Toward an interpretable Alzheimer's disease diagnostic model with regional abnormality representation via deep learning.* Neuroimage 2019; 202:116113.
102. Papanastasopoulos Z, Samala RK, Chan HP, et al. *Explainable AI for medical imaging: deep-learning CNN ensemble for classification of estrogen receptor status from breast MRI.* SPIE Medical Imaging, vol. 11314. SPIE; 2020.
103. Wu YH, Gao SH, Mei J, et al. JCS: *an explainable COVID-19 diagnosis system by joint classification and segmentation.* IEEE Trans Image Process 2021;30: 3113–26.
104. Gunraj, H., L. Wang, and A. Wong, *COVIDNet-CT: A Tailored Deep Convolutional Neural Network Design for Detection of COVID-19 Cases From Chest CT Images.* Front Med (Lausanne), 2020. 7: p. 608525.
105. Monga, V., Y. Li, and Y.C. Eldar, *Algorithm Unrolling: Interpretable, Efficient Deep Learning for Signal and Image Processing.* IEEE Signal Processing Magazine, 2021. 38(2): p. 18-44.
106. Li Y, Tofighi M, Geng J, et al. *Efficient and interpretable deep blind image deblurring via algorithm unrolling.* IEEE Trans Comput Imaging 2020; 6:666–81.
107. Marivani I, Tsiligianni E, Cornelis B, et al. *Multimodal deep unfolding for guided image super-resolution.* IEEE Trans Image Process 2020; 29:8443–56.
108. Huang Y, Li J, Gao X, et al. *Interpretable detail-fidelity attention network for single image super-resolution.* IEEE Trans Image Process 2021; 30:2325–39.
109. Yokota T, Hontani H, Zhao Q, et al. *Manifold modeling in embedded space: an interpretable alternative to deep image prior.* IEEE Trans Neural Netw Learn Syst. 2020. p. 1-15